\documentstyle[12pt]{article}
\input amssym.def
\input amssym

\def\cc{\circ_\nu}
\def\aa{{\scriptstyle\odot}_\nu}
\def\la{\lambda}

\newtheorem{theorem}{Theorem}
\newtheorem{proposition}{Proposition}
\newtheorem{definition}{Definition}
\newtheorem{lemma}{Lemma}

\newtheorem{corollary}{Corollary}
\newenvironment{proof}{\noindent{\it Proof.\/}}{\vskip3mm}
\newenvironment{acknowledgement}{{\vskip3mm}
{\noindent{\bf Acknowledgements.\/}}}{\vskip3mm}

\begin{document}
\title{Generalized Abelian Deformations: Application to Nambu Mechanics}
\author{Giuseppe Dito${}^{1}$\thanks{Supported by the European
       Commission and the Japan Society for the Promotion of Science.}
       \ and Mosh\'e Flato${}^{2}$\\ \\
       ${}^1$ Research Institute for Mathematical Sciences\\ Kyoto
       University\\ Kitashirakawa, Oiwake-cho\\ Sakyo-ku, Kyoto 606-01
       Japan\\ \\ ${}^2$ D\'epartement de Math\'ematiques\\ Universit\'e de
       Bourgogne\\ BP 138, F-21004 Dijon Cedex France\\}
\date{September 1996}
\maketitle
\vspace*{-2mm}\noindent{\bf Abstract:}
We study Abelian generalized deformations of the usual product of
polynomials introduced in \cite{DFST}. We construct an explicit
example   for the case of $\frak{su}(2)$ which provides a
tentative of a quantum-mechanical
description of Nambu Mechanics on ${\Bbb R}^3$.
By introducing the notions of strong and weak triviality
of generalized deformations, we show that the Zariski product is never
trivial in either sense, while the example constructed here in a
quantum-mechanical context is only strongly non-trivial.
\thispagestyle{empty}
\newpage\setcounter{page}{1}
\section{Introduction}
The generalization of Hamiltonian Mechanics introduced more than 20
years ago by Nambu \cite{Na} has been recently formulated within a geometrical
framework \cite{Ta}. The importance of the so-called Fundamental
Identity (FI) \cite{FF,Ta} has been recognized as a dynamical
consistency condition for Nambu's original formulation. Somehow the FI
plays the r\^ole of the Jacobi identity for Poisson bracket in the
usual Hamiltonian Mechanics, but its consequences are by far
completely different as it imposes strong constraints on the
underlying geometrical structure \cite{Ta}. The reader is referred to
\cite{Ta} for further details on Nambu structures and
manifolds, we shall limit ourselves to recall the definition.
Let $M$ be a $m$-dimensional manifold. Denote by $N$ the algebra of
smooth real-valued functions on $M$. A Nambu bracket of order $n$
on $M$ is a $n$-linear map on $N$ taking values in $N$, denoted
by $(f_1,\ldots,f_n)\mapsto \{f_1,\ldots,f_n\}$, $f_i \in N$, such that
the following properties are satisfied for any functions
$f_0,\ldots,f_{2n-1}\in N$:\vskip3mm
\noindent a) Skew-symmetry
$$
\{f_1,\ldots,f_n\} = \epsilon(\sigma)
\{f_{\sigma_1},\ldots,f_{\sigma_n}\}, \quad \forall \sigma\in S_n;
$$
b) Leibniz rule
$$
\{f_0f_1,f_2,\ldots,f_n\} =
f_0\{f_1,f_2,\ldots,f_n\}+\{f_0,f_2,\ldots,f_n\}f_1 ;
$$
c) Fundamental Identity
\begin{eqnarray*}
\{f_1,\ldots,f_{n-1},\{f_{n},\ldots,f_{2n-1}\}\}
&=&\{\{f_1,\ldots,f_{n-1},f_n\},f_{n+1},\ldots,f_{2n-1}\}\\
&&\quad +\{f_n,\{f_1,\ldots,f_{n-1},f_{n+1}\},f_{n+2},\ldots,f_{2n-1}\}\\
&&\quad +\cdots+\{f_{n},f_{n+1},\ldots,f_{2n-2},
\{f_1,\ldots,f_{n-1},f_{2n-1}\}\};
\end{eqnarray*}
where $S_n$ is the group of permutations of the set $\{1,\ldots,n\}$
and $\epsilon(\sigma)$ is the sign of the permutation $\sigma\in S_n$.
A Nambu bracket defines a Nambu structure on $M$. Then $M$ is said to
be a Nambu-Poisson manifold.

The quantization of Nambu structures turns out to be a non-trivial problem,
even (or especially) in the simplest cases. Usual approaches to quantization
have failed to give an appropriate solution, and in a common work
\cite{DFST} with D.~Sternheimer and L.~Takhtajan, we were led to
introduce a new quantization scheme (Zariski quantization) in order to
give a solution to that old problem. The central idea of Zariski
quantization is to look first for an Abelian deformation of the usual
product of functions instead of a direct deformation of the Nambu bracket.
Then the quantization of the Nambu bracket is achieved by plugging in
the Abelian deformed product into the classical Nambu bracket.

Abelian algebra deformations are classified according to the
Harrison cohomology, and  the second Harrison  cohomology group is
trivial for the space of polynomials on ${\Bbb R}^n$.
One way to overcome this cohomological
difficulty is to consider
generalized deformations not of the usual Gerstenhaber-type.
In the case of Zariski quantization, linearity
with respect to the deformation parameter does not hold:
The deformed product operation
annihilates the deformation parameter, so the usual Harrison
cohomology is not applicable. This product
is obtained by factorization of real polynomials into irreducible
factors and complete symmetrization by a partial Moyal product.
However the deformed product is not distributive  when considered
as a product between polynomials (it is only Abelian
and associative). The linearization of this deformed product
(by extending it to a semi-group algebra), called Zariski product,
provides an Abelian algebra deformation of the usual product
on the semi-group algebra generated by
the set of real irreducible polynomials.
By an appropriate presentation of this algebra which allows a coherent
notion of derivatives, we can define in a natural way deformation of
the classical Nambu bracket and, hence, a quantization of the usual
Nambu structure on the semi-group algebra.

The Zariski quantization looks like a field quantization where irreducible
polynomials are considered as one-particle states. One may wonder
if, starting from the Moyal product, a quantum-mechanical approach
would have been  possible if one
had made a decomposition into linear factors instead of
irreducible ones. The answer is negative: Complete symmetrization by
Moyal product of linear factors provides no quantization at all.
However, as indicated in \cite{DFST}, the situation is completely
different if one considers an invariant star-product on
$\frak{su}(2)^* \sim {\Bbb R}^3$ (the one which is associated
with the quantization of angular momentum, the linear factors
being here the generators of the Lie algebra).
In that case it is possible to find a quantization by staying in the
algebra of smooth functions on ${\Bbb R}^3$.

In this paper we shall work out some remarks
stated in \cite{DFST} regarding
a possible  quantum-mechanical approach (with finite number of degrees
of freedom), non-triviality of the generalized deformations (i.e. not of
Gerstenhaber-type) and develop the example for $\frak{su}(2)^* $.
It is shown in Section~2 that some kinds of ${\Bbb R}[\nu]$-linear
deformations are not interesting in our context, in the sense
that they cannot be associative without coinciding with the usual product.
In Section~3, we shall make precise what we call a quantum-mechanical
approach to generalized Abelian deformations.
The questions of non-triviality and equivalence of these deformations are
also studied. We introduce the notions of A- and B-equivalences
(and the corresponding notions of triviality: strong and weak) for
generalized deformations, and in particular we show that the Zariski product
is non-trivial in both senses (weak and strong).
A detailed example is presented in Section~4
for the $\frak{su}(2)^* $-case by giving an explicit form of the deformed
product in terms of differential operators, providing a strongly
non-trivial deformation of the usual product on ${\Bbb R}^3$.
We conclude this paper by
some remarks on some spectrality properties of the generalized Abelian
deformations and their application to the quantization of
the Nambu bracket.
\section{${\Bbb R}[\nu]$-linear products}
The generalized deformation introduced in \cite{DFST} is not an
algebra deformation in the sense of Gerstenhaber. Also, when
restricted to the space of polynomials on ${\Bbb R}^n$, it is not
additive due to the factorization into irreducible polynomials
involved in the definition of the product. If instead we had performed a
decomposition (by addition of monomials and multiplicative
factorization) into {\it linear\/} polynomials, it would have
been possible to stay in the framework of Gerstenhaber-type
deformations by requiring linearity with respect to the
deformation parameter. However, we shall show here that this attempt
is not interesting: First, we would have found nothing but a trivial
deformation of the usual product (the Harrison cohomology is trivial
here) and, which is  worst, one cannot expect to conciliate
associativity of this product with quantization. We shall show
that this kind of products are associative if and only if they
coincide with the usual product of functions.

Consider ${\Bbb R}^n$ with coordinates denoted by $(x_1,\ldots,x_n)$.
Let $P$ be a Poisson bracket on ${\Bbb R}^n$ and
$\ast$ a star-product on ${\Bbb R}^n$ with deformation parameter
$\nu$ such that the star-bracket $[f,g]_\ast \equiv (f\ast g - g\ast
f)/2\nu$, $f,g \in C^\infty({\Bbb R}^n)$, defines a Lie algebra
deformation of the Lie algebra $(C^\infty({\Bbb R}^n),P)$
(see \cite{BFFLSI,BFFLSII} for the first extensive papers on star-products).
Denote by $N$ the algebra of real polynomials on ${\Bbb R}^n$ and
by $N[\nu]$ the space of polynomials in $\nu$ with
coefficients in $N$. The symmetric tensor algebra of $N$ (with scalars)
is denoted by ${\cal S}$ with symmetric tensor product $\otimes$.
${\cal S}[\nu]$ is the space of polynomials in $\nu$ with
coefficients in ${\cal S}$.

Following the lines of \cite{DFST}, we define a ${\Bbb R}[\nu]$-linear
map $\la\colon N[\nu]\rightarrow {\cal S}[\nu]$ by
\begin{equation}\label{defla}
\la(x_1^{k_1}\cdots x_n^{k_n})
=(x_1^{k_1\atop\otimes})\otimes\cdots\otimes (x_n^{k_n\atop\otimes}),
\quad \forall k_1,\cdots,k_n \geq 0.
\end{equation}
In particular, $\la(1)=I$, the identity of ${\cal S}[\nu]$.

Let $T\colon {\cal S}[\nu]\rightarrow N[\nu]$ be the unique
${\Bbb R}[\nu]$-linear map defined by $T(I)=1$ and
\begin{equation}\label{defT}
T(F_1\otimes\cdots\otimes F_k)={1\over k!}\sum_{\sigma\in S_k}
F_{\sigma_1}\ast\cdots\ast F_{\sigma_k},\quad \forall k\geq 1,
\end{equation}
where $F_i\in N$, $1\leq i\leq k$.

\begin{definition}
Let $\ast$ be a star-product on ${\Bbb R}^n$ endowed with some Poisson
bracket. Let us define a new product $\cc$ on $N[\nu]$ by the
following formula:
\begin{equation}\label{defcc}
F\cc G = T(\la(F)\otimes \la(G)), \quad \forall F, G \in N[\nu].
\end{equation}
It is a ${\Bbb R}[\nu]$-distributive  Abelian product. We shall call it
the $\cc$-product associated with $\ast$.
\end{definition}
In general, the product defined by Eq.~(\ref{defcc}) is not
associative, and the following shows that it is associative only
in the trivial case:
\begin{proposition}
Let $\ast$ be a star-product on ${\Bbb R}^n$ such that its associated
$\cc$-product is associative, then the product $\cc$ is the usual
pointwise product on $N$.
\end{proposition}
\begin{proof}
It is easy to see that the map $\la$ is an algebra homomorphism:
$\la(FG)=\la(F)\otimes\la(G)$, $\forall F, G \in N[\nu]$. Hence
the product $F\cc G$, $F, G \in N[\nu]$,  defined by
Eq.~(\ref{defcc}) can be written in the following form:
\begin{equation}\label{homo}
F\cc G =T(\la(FG))=\sum_{r\geq0} \nu^r \rho_r (FG),
\end{equation}
where $\rho_0(FG)=FG$ and $\rho_r$, $r\geq1$, are ${\Bbb R}[\nu]$-linear
maps on $N[\nu]$ (whose restrictions to $N$ are ${\Bbb R}$-linear map on $N$).

Using Eq.~(\ref{homo}), the associativity of the product $\cc$ can
be formulated as
\begin{equation}\label{asso}
\sum_{r\geq0}\nu^r\sum_{u+v=r\atop u,v\geq0}\rho_u(\rho_v(FG)H)
=\sum_{r\geq0}\nu^r\sum_{u+v=r\atop u,v\geq0}\rho_u(F\rho_v(GH)),
\quad\forall F,G,H\in N[\nu].
\end{equation}
Consider the last equation for $F,G,H\in N$. By identifying
the coefficients of the different powers of $\nu$ on both side, we find
for $r=0$: FGH=FGH, and for $r=1$:
\begin{equation}\label{asso1}
\rho_1(FG)H = F\rho_1(GH), \quad \forall F,G,H\in N.
\end{equation}
Set $G=H=1$ in Eq.~(\ref{asso1}), one finds $\rho_1(F)=F\rho_1(1)$,
$\forall F\in N$. By definition, $T(\la(1))=1$ and Eq.~(\ref{homo})
implies $\rho_r(1)=0$, $\forall r\geq 1$, thus $\rho_1(F)=0$,
$\forall F\in N$.

Now suppose that for some $k \geq 2$ we have $\rho_i(F)=0$ for all
$1\leq i \leq k$, $\forall F\in N$. By equating the coefficients of
$\nu^{k+1}$ on both sides of Eq.~(\ref{asso}), we find that
$$
\rho_{k+1}(FG)H = F\rho_{k+1}(GH), \quad \forall F,G,H\in N,
$$
and by the same argument used for showing that $\rho_1(F)=0$,
$\forall F\in N$, we find $\rho_{k+1}(F)=0$, $\forall F\in N$.
In conclusion, $\rho_{r}(F)=0$, $\forall r\geq1$, $\forall F\in N$,
and Eq.~(\ref{homo}) gives $F\cc G = FG$, $\forall F,G\in N$ and this
shows the Proposition.
\end{proof}
\section{Non-Triviality and Equivalence}
In Zariski quantization, the map $\alpha\colon N[\nu]\rightarrow
{\cal S}$ which replaces the usual product
appearing in the decomposition of
some polynomial into irreducible factors by the symmetric tensor
product, is not ${\Bbb R}[\nu]$-multiplicative \cite{DFST}. In order
to keep  associativity, the map $\alpha$ has to annihilate non-zero
powers of $\nu$.  Also the space $N[\nu]$ endowed with the Zariski
product is not an Abelian algebra, it is only an Abelian semi-group.
Indeed the distributivity with respect to the addition of $N$ is lacking due
to the factorization into irreducible factors. However if one replaces
``factorization into irreducible factors'' by ``factorization
of monomials into linear factors (and linear combinations)''
in the definition of Zariski products, then one does
get an Abelian algebra structure on $N[\nu]$. In the case of the Moyal
product, for reasons explained in the Introduction, this does not lead
to a new product on $N[\nu]$, and one has to go through the
construction developed in \cite{DFST}.
However there are other star-products than Moyal
and in general it is possible to find a
generalized Abelian algebra deformation in a quantum-mechanical
framework \cite{Di}. The example of $\frak{su}(2)$ will be treated
to some extend in Section~4
(in this case the linear factors are generators of the Lie algebra
$\frak{su}(2)$, linear functions on $\frak{su}(2)^* \sim {\Bbb R}^3$;
they have a quadratic expression in ${\Bbb R}^6$).

After making precise what we call a quantum-mechanical approach
by defining a deformed Abelian associative product on $N[\nu]$,
we shall be concerned with the possible definitions of equivalence and
triviality of Abelian generalized deformations.
Also we shall discuss their consequences for both
$\aa$-products studied in the present paper and Zariski products
introduced in \cite{DFST}.

Let us first define an Abelian product in a
way similar to what is done in \cite{DFST},
with the main difference that it is ${\Bbb R}$-distributive.
Let $\pi\colon N[\nu]\rightarrow N$ be the natural projection
of $N[\nu]$
onto $N$. Let $\la_0 =\la \circ \pi\colon N[\nu]\rightarrow {\cal S}$,
where $\la$ is defined by Eq.~(\ref{defla}), and let
$\tilde T\colon{\cal S}\rightarrow N[\nu]$ be the restriction to
${\cal S}$ of the map defined by Eq.~(\ref{defT}).
\begin{definition}
Let $\ast$ be a star-product on ${\Bbb R}^n$. We shall call the product $\aa$
on $N[\nu]$ defined by
\begin{equation}\label{defaa}
F\aa G = \tilde T(\la_0(F)\otimes \la_0(G)), \quad \forall F, G \in N[\nu],
\end{equation}
the $\aa$-product associated with $\ast$.
\end{definition}
Contrary to the $\circ_\nu$-products of Section~2, the $\aa$-products
annihilate all non-zero powers of the deformation parameter $\nu$.
Clearly the product $\aa$ is Abelian and distributive with respect to
the addition in $N[\nu]$, but is not ${\Bbb R}[\nu]$-multiplicative,
i.e., in general $(aF)\aa G \neq a (F\aa G)$ for $a\in{\Bbb R}[\nu]$.
However, it is always associative as shown by:
\begin{lemma}
Let $\ast$ be a star-product on ${\Bbb R}^n$. Its associated
$\aa$-product is associative.
\end{lemma}
\begin{proof}
Simply notice that $\la_0(F\aa G)=\la_0(F)\otimes\la_0(G)$, $\forall
F,G\in N[\nu]$,  and thus
\begin{eqnarray*}
&F\aa (G\aa H)&=\tilde T(\la_0(F)\otimes\la_0(G\aa H))=
\tilde T(\la_0(F)\otimes\la_0(G)\otimes\la_0(H))\\
&{}&=\tilde T(\la_0(F\aa G)\otimes \la_0(H))=(F\aa G)\aa H,\quad \forall
F,G,H\in N[\nu].
\end{eqnarray*}
\end{proof}
Hence $N[\nu]$ endowed with a $\aa$-product defines an Abelian
associative generalized deformation of the usual product on $N$.
This makes $(N[\nu],\aa)$ into an Abelian ${\Bbb R}$-algebra. Let us mention
that in general a $\aa$-product associated with some star-product
differs from the usual product on $N$, and the star-products which
have the usual product as associated  $\aa$-product are all of the
Moyal-type \cite{Di}.
More precisely, on the dual of a Lie algebra, there exists one and
only one covariant star-product whose associated $\aa$-product is the
usual product of polynomials.

We shall now be concerned with the question of equivalence of
$\aa$-products. Several notions of equivalence can be adapted to the
kind of generalized deformations considered in the present paper.
Let us present a first definition of equivalence which is similar to
the usual notion of equivalence for Gerstenhaber-type deformations
for associative algebras. First notice that
since the map $\lambda_0$ is a homomorphism and it annihilates
non-zero powers of the parameter $\nu$, it is easy to find that
any $\aa$-product can be written in the form
\begin{equation}\label{dege}
F\aa G = FG + \sum_{r\geq 0} \nu^r \rho_r (F G),\quad F,G \in N,
\end{equation}
where $\rho_r\colon N \rightarrow N$ are linear maps (the `cochains'
of the $\aa$-product). As for associative deformations, one way to
understand equivalence of $\aa$-products is given by the following:
\begin{definition}
Two products $\aa$ and $\aa'$ are said to be A-equivalent, if there exists a
${\Bbb R}[\nu]$-linear (formally invertible) map $S_\nu\colon
N[[\nu]]\mapsto N[[\nu]]$ of the form $S_\nu=\sum_{r\geq 0} \nu^r S_r$,
where $S_r\colon N\mapsto N$, $r\geq1$, are differential operators
and $S_0=Id$, such that
$$ \left. S_\nu(F \aa G) = S_\mu (F) \aa' S_\mu (G)\right|_{\mu= \nu},
\quad F,G \in N, $$
which amounts to
$$
\sum_{r,s\geq0} \nu^{r+s}S_s(\rho_r(FG))
= \sum_{r,s,s'\geq0}\nu^{r+s+s'}\rho_r'(S_s(F)S_{s'}(G)),\quad F,G \in N,
$$
where $\rho_r$ (resp. $\rho_r'$) are the cochains of the product
$\aa$ (resp. $\aa'$).
\end{definition}
It is straightforward to check that
Definition~3 indeed defines an equivalence relation between
$\aa$-products.
One has the associated notion of triviality given by:
\begin{definition}
A $\aa$-product associated with some star-product is said to be strongly
trivial if it is A-equivalent to the usual product, i.e.,
there exists a ${\Bbb R}[\nu]$-linear (formally invertible) map
$S_\nu\colon N[[\nu]]\mapsto
N[[\nu]]$ of the form $S_\nu=\sum_{r\geq 0} \nu^r S_r$, where
$S_r\colon N\mapsto N$, $r\geq1$, are differential operators
and $S_0=Id$, such that:
\begin{equation}\label{trivial}
S_\nu(F\aa G) = S_\nu(F)\cdot S_\nu(G), \quad \forall F,G\in N,
\end{equation}
where $\cdot$ denotes the usual product.
\end{definition}

The following shows that in general a $\aa$-product is strongly non-trivial.
\begin{proposition}
A $\aa$-product is strongly trivial if and only if it coincides with the
usual product.
\end{proposition}
\begin{proof}
Let $\aa$ be strongly trivial.  Since $1\aa1 =1$, the map $S_\nu$ in
(\ref{trivial}) must satisfy $S_\nu(1)=S_\nu(1)\cdot S_\nu(1)$, which
implies $S_\nu(1)=1$, i.e., $S_r(1) =0$, $\forall r\geq 1$.

The map $\lambda_0$ in (\ref{defaa}) is a homomorphism, it implies that
the product $F\aa G$, $F,G\in N$, can be expressed as
$$
F\aa G =\sum_{r\geq0} \nu^r \rho_r (FG),\quad F,G\in N,
$$
where $\rho_0(FG)=FG$ and $\rho_r$, $r\geq1$, are linear
maps on $N$. By substituting this last expression for $F\aa G$
in Eq.~(\ref{trivial}), and by identifying the different powers of
$\nu$ on both sides, we find:
\begin{equation}\label{aaa}
\sum_{r+s=t\atop r,s\geq0}S_r(\rho_s(FG)) =
\sum_{r+s=t\atop r,s\geq0}S_r(F) S_s(G),
\quad \forall t\geq 0, F,G\in N.
\end{equation}
Now set $G=1$ in the preceding equality. Since $S_r(1)=0$ for
$r\geq1$, Eq.~(\ref{aaa}) yields
\begin{equation} \label{aaa1}
S_t(F) = \sum_{r+s=t\atop r,s\geq0}S_r(\rho_s(F)),
\quad \forall t\geq0,F\in N.
\end{equation}
For $t=1$, we find that $\rho_1(F)=0$, $\forall F\in N$. For $t\geq2$,
Eq.~(\ref{aaa1}) can be written in the following form:
$$
S_t(F) = S_t(F) + \sum_{r+s=t\atop r,s\geq1}S_r(\rho_s(F)) + \rho_t(F)
\quad \forall t\geq2,F\in N,
$$
i.e, $\rho_t(F)=-\sum_{r+s=t\atop r,s\geq1}S_r(\rho_s(F))$,
$\forall t\geq2,F\in N$. Since $\rho_1(F)=0$, $\forall F\in N$, it
is easy to show by induction that
$\rho_t(F)=0$, $\forall t\geq1,F\in N$. Therefore, if $\aa$ is
strongly trivial, it coincides with the usual product. The converse
statement being obvious, the Proposition holds.
\end{proof}

Actually the treatment done above applies almost
straightforwardly to the case of Zariski quantization yielding
that the Zariski product is never strongly trivial.
Let us denote by $\bullet_\nu$ the Zariski product constructed out from
some star-product which is defined
on the algebra ${\cal A}_\nu$ (the reader is referred to \cite{DFST}
for the definitions of  $\bullet_\nu$ and ${\cal A}_\nu$;
the deformation parameter $\nu$ was taken to be $\hbar$ in the preceding
reference). The product $\bullet_\nu$ has also the form given by
Eq.~(\ref{dege}) where now the maps $\rho_r$ are linear maps on
${\cal A}_0$, the classical algebra, and the ``usual'' product
has to be interpreted as
the product on the algebra ${\cal A}_0$. Recall that
there are derivations defined on ${\cal A}_\nu$ \cite{DFST} and hence we have a
natural definition of ``differential operators'' acting on ${\cal A}_\nu$.
By an obvious adaptation of Definitions~3 and 4, one gets the corresponding
definitions of A-equivalence and strong triviality for Zariski products.
Notice that the proof of Proposition~2 applies literally to the case of
Zariski products (the differential aspect of the intertwining operator
is not involved in the proof). However, by construction,
a Zariski product constructed out from some star-product can never
coincide with the ``usual product'' on ${\cal A}_0$ (of course, except in the
trivial situation where  the  star-product is the usual product of
polynomials on ${\Bbb R}^n$). This leads to:
\begin{theorem}
Let $\bullet_\nu$ be a Zariski product associated with some
star-product $\ast_\nu$ on ${\Bbb R}^n$, then $\bullet_\nu$ is
strongly non-trivial whenever $\ast_\nu$ is not the usual product.
\end{theorem}
Another natural definition of equivalence
for the generalized deformations that one can consider is given
by the following:
\begin{definition}
Two products $\aa$ and $\aa'$ are said to be B-equivalent,
if there exists a ${\Bbb R}[\nu]$-linear (formally invertible) map $S_\nu\colon
N[[\nu]]\mapsto N[[\nu]]$ of the form $S_\nu=\sum_{r\geq 0} \nu^r S_r$,
where $S_r\colon N\mapsto N$, $r\geq1$, are differential operators
and $S_0=Id$, such that:
$$
S_\nu(F\aa G)= S_\nu(F)\aa' S_\nu(G), \quad F,G \in N,
$$
\end{definition}
Since the $\aa$-product annihilates the parameter $\nu$, the preceding
definition is equivalent to say that $S_\nu(F\aa G)=
F\aa'G$, $F,G \in N$, holds. The difference between this
definition of equivalence and that given by Definition~3 lies in the fact
the latter is formulated in such a way that the terms involving the
powers of the parameter $\nu$ introduced by the intertwining operator
$S_\nu$ are taken into account, while the former definition allows the
product operation to annihilate them.  Also notice that two
$\aa$-products which are B-equivalent are not necessarily A-equivalent.

The notion of triviality corresponding to B-equivalence (weaker than strong
triviality, cf. Proposition~3, hence the choice of terminology) is:

\begin{definition}
A $\aa$-product associated with some star-product is said to be
weakly trivial if it is B-equivalent to the usual product, i.e.,
there exists a ${\Bbb R}[\nu]$-linear (formally invertible) map
$S_\nu\colon N[[\nu]]\mapsto
N[[\nu]]$ of the form $S_\nu=\sum_{r\geq 0} \nu^r S_r$, where
$S_r\colon N\mapsto N$, $r\geq1$, are differential operators
and $S_0=Id$, such that:
$$
S_\nu(F\cdot G) = F \aa G, \quad \forall F,G\in N,
$$
where $\cdot$ denotes the usual product.
\end{definition}
By adequate modifications, these definitions also apply  to the
case of Zariski products.

Let us study the consequences of these definitions for the
$\aa$-products and then for the Zariski products.
Consider a $\aa$-product whose cochains are given by differential operators
(e.g. the $\aa$-product of Section~4). Then the $\aa$-product has
the form $F\aa G = \rho(\pi(F\cdot G))$, $F,G\in N[\nu]$, where $\rho=Id +
\sum_{r\geq1} \nu^r \rho_r$ and the $\rho_r$'s are differential
operators acting on $N$, and $\pi\colon N[\nu]\mapsto N$ is the
projection onto the classical part.
Note that $\rho$ is formally invertible, and we have $\rho(F\cdot G)= F\aa G$,
$F,G\in N$, therefore  $\rho$ intertwines the $\aa$-product with the usual
product in the sense of Definition~6. Hence we have shown the following:
\begin{proposition}
A $\aa$-product is weakly trivial whenever its cochains are given by
differential operators.
\end{proposition}
For a Zariski product $\bullet_\nu$, it is true that the corresponding $\rho$
intertwines  $\bullet_\nu$ with the usual product on ${\cal A}_0$, but
the cochains $\rho_r$ are never given by differential operators
acting on ${\cal A}_0$.
Hence
\begin{corollary}
A Zariski product is weakly non-trivial (except
when the defining star-product is the usual product).
\end{corollary}
\vskip1mm
\noindent{\bf Remark 1} One may wonder why we restrict  the
intertwining operators to be defined by
{\it differentiable} cochains as in general the deformations considered
are not defined by differentiable cochains. This is so because eventually the
non-triviality for the Abelian generalized deformations
should be linked with the usual (differentiable) Hochschild or Harrison
cohomologies.
\vskip1mm
\noindent{\bf Remark 2}
Of course the map $T\circ\alpha$ which defines
the Zariski product satisfies
$$T\circ\alpha(F)\bullet_\nu T\circ\alpha(G)=F\bullet_\nu G=
T\circ\alpha(F\cdot G), \quad\forall F,G \in {\cal A}_0,$$
where $\cdot$
denotes the product on the classical algebra ${\cal A}_0$. But the map
$T\circ\alpha$ is neither invertible as a formal series nor acting by
differential operators. Hence the map $T\circ\alpha$ is not an
intertwiner trivializing the Zariski product in the
sense of Definition~6.
Even if one defines $S_\nu$ by taking the ${\Bbb R}[[\nu]]$-linear
extension of the restriction of $T\circ\alpha$ to ${\cal A}_0$,
one gets an invertible map but not an intertwining map (in the sense of
B-equivalence) as $S_\nu$ is not given by differential operators
on ${\cal A}_0$.
\vskip1mm

To sum up, the kind of Abelian deformations defined by a Zariski
product considered in~\cite{DFST} are
both strongly and weakly non-trivial. However the example developed
in Section~4 is strongly non-trivial but weakly trivial.
We think that the above considerations may give some hints for the
definition of an appropriate cohomology for Abelian generalized
deformations.
\section{Example of $\frak{su}(2)$}
Here we shall study in some details an example of $\aa$-product
associated with an invariant star-product on the dual of the
$\frak{su}(2)$-Lie algebra ($\sim {\Bbb R}^3$). This $\aa$-product
is strongly non-trivial in the sense of Definitions~3 and 4.
An explicit formula for that $\aa$-product will be given,
which permits to extend the $\aa$-product to $C^{\infty}({\Bbb R}^3)$
(valued in $C^{\infty}({\Bbb R}^3)[[\nu]]$, whereto it extends trivially
since the $\aa$-product annihilates $\nu$).
Let $\ast^{}_M$ be the Moyal product on ${\Bbb R}^6$:
$$
F\ast^{}_M G = \sum_{r\geq0} \nu^r P_M^{(r)}(F,G), \quad \forall
F,G\in C^\infty({\Bbb R}^6),
$$
where $P_M^{(r)}$ is the $r$-th power of the Poisson bracket:
$$
P_M^{}(F,G) = \sum_{1\leq i\leq 3} \Bigl( {\partial F \over\partial p_i }
{\partial G \over\partial q_i } -
{\partial F \over\partial q_i }
{\partial G \over\partial p_i }\Bigr).
$$
Consider the functions $L_i(p,q) = \sum_{1\leq j,k \leq 3}
\varepsilon_{ijk} p_j q_k$, $1\leq i\leq 3$, on ${\Bbb R}^6$, where
$\varepsilon_{ijk}$ is the totally skew-symmetric tensor with
$\varepsilon_{123}=1$. The $L_i$'s realize the Lie algebra $\frak{su}(2)$
for  the Moyal bracket $[\cdot,\cdot]_M^{}$ on ${\Bbb R}^6$, i.e.,
$$
 [L_i,L_j]_M^{}\equiv
{L_i\ast^{}_M L_j  - L_j \ast^{}_M L_i\over2\nu}
=  \sum_{1\leq k \leq 3}\varepsilon_{ijk} L_k,\quad 1\leq i,j\leq 3.
$$
One can easily show that for any polynomial
$F\colon{\Bbb R}^3\rightarrow{\Bbb R}$, the function $F(L_1,L_2,L_3)$
on ${\Bbb R}^6$
satisfies
\begin{equation}
\label{lproduct}
L_i\ast^{}_M F= L_i F +\nu\sum_{1\leq j,k \leq 3}
\varepsilon_{ijk} L_k {\partial F \over\partial L_j }
+ \nu^2 \Bigl( 2 {\partial F \over\partial L_i} +\sum_{1\leq j \leq 3}
L_j {\partial^2 F \over\partial L_i \partial L_j }\Bigr), \quad 1\leq i
\leq 3.
\end{equation}
We observe that $L_i\ast^{}_M F(L_1,L_2,L_3)$ is still a polynomial in the
variables $(L_1,L_2,L_3)$. By induction on $k$, it is readily shown
that $L_{i_1}\ast^{}_M\cdots\ast^{}_M L_{i_k}$ is a polynomial in
$(L_1,L_2,L_3)$. Also any polynomial in $(L_1,L_2,L_3)$ can be
expressed as a $\ast^{}_M$-polynomial of the $L_i$'s, so the product
$F \ast^{}_M G$ of two polynomials $F,G$ in $(L_1,L_2,L_3)$ is
a polynomial in $(L_1,L_2,L_3)$. Hence from the Moyal product
$\ast^{}_M$ on ${\Bbb R}^6$, we get a star-product on ${\Bbb R}^3$ satisfying
Eq.~(\ref{lproduct}) for any polynomial $F$. This star-product, that
we shall denote by $\ast$, is actually an invariant (and
covariant) star-product on $\frak{su}(2)^* \sim {\Bbb R}^3$.

Let $\aa$ be the $\aa$-product associated with $\ast$. Remark that
$$
L_i\ast L_j = L_i L_j +\nu\sum_{1\leq k \leq 3}\varepsilon_{ijk} L_k
+ 2 \nu^2 \delta_{ij},\quad 1\leq i,j \leq 3,
$$
and $L_i\aa L_j = L_i L_j + 2 \nu^2 \delta_{ij}$, so this
$\ast$-product does provide quantum terms.
In the following we shall derive an explicit expression for
$\aa$. Before stating the main
result of this section, we need some combinatorial
preliminaries. We shall denote the cochains
of the star-product $\ast$ on ${\Bbb R}^3$ by $C_r$, i.e., $F\ast G =
\sum_{r\geq 0} \nu^r C_r(F,G)$, where $C_1(F,G)= P(F,G)\equiv
\sum_{1\leq i,j,k \leq 3}\varepsilon_{ijk} L_k
{\partial F \over\partial L_i}
{\partial G \over\partial L_j}$ is the standard Poisson bracket on
$\frak{su}(2)^* \sim {\Bbb R}^3$.
We denote by $\Delta$ the Laplacian
operator: $\Delta=\sum_{1\leq k\leq3} {\partial^2\over\partial L_k^2}$.
\begin{lemma}
For any $n\geq 1$ and any indices $i_1,\ldots,i_n\in\{1,2,3\}$, we have:
\begin{eqnarray*}
&{\rm a)}& \sum_{(i_1,\ldots,i_n)}{\partial\over\partial L_{i_1}}
(L_{i_2}\cdots L_{i_n}) = \Delta (L_{i_1}\cdots L_{i_n});\cr
 & & \cr
&{\rm b)}& \sum_{(i_1,\ldots,i_n)}L_{i_1}\Delta^m (L_{i_2}\cdots
  L_{i_n}) = (n-2m)\Delta^m (L_{i_1}\cdots L_{i_n}),
\quad \forall m\geq0;\cr
 & & \cr
&{\rm c)}& \sum_{(i_1,\ldots,i_n)}P(L_{i_1},
\Delta^m (L_{i_2}\cdots L_{i_n}))= 0,
\quad \forall m\geq0;\cr
 & & \cr
&{\rm d)}& \sum_{(i_1,\ldots,i_n)}C_2(L_{i_1},\Delta^m (L_{i_2}\cdots
  L_{i_n}))= (n-2m)\Delta^{m+1} (L_{i_1}\cdots L_{i_n}),
\quad \forall m\geq0;
\end{eqnarray*}
where $\displaystyle\sum_{(i_1,\ldots,i_n)}$ stands for summation over cyclic
permutations of $(i_1,\ldots,i_n)$, and we set $L_{i_2}\cdots
L_{i_n}=1$ when $n=1$.
\end{lemma}
\begin{proof}
Statement a) is proved by performing the derivations with respect to
the $L_i$'s on both side. We have that
$$
{\partial\over\partial L_{i_1}}
(L_{i_2}\cdots L_{i_n})=\sum_{2\leq k\leq n} \delta_{i_1 i_k}
L_{i_2}\cdots\hat{L}_{i_k}\cdots L_{i_n},
$$
where the symbol $\hat{ }$ stands for omission. The sum on the
right-hand side can be written as $\sum_{1\leq k\leq n\atop k\neq1}
\delta_{i_1 i_k} \hat{L}_{i_1}L_{i_2}\cdots\hat{L}_{i_k}\cdots
L_{i_n}$. The sum over cyclic permutations gives
\begin{eqnarray*}
\sum_{(i_1,\ldots,i_n)}{\partial\over\partial L_{i_1}}
(L_{i_2}\cdots L_{i_n}) & = & \sum_{(i_1,\ldots,i_n)}
\sum_{1\leq k\leq n\atop k\neq1}
\delta_{i_1 i_k} \hat{L}_{i_1}L_{i_2}\cdots\hat{L}_{i_k}\cdots
L_{i_n}\cr
& = & \sum_{1\leq s\leq n}\sum_{1\leq k\leq n\atop k\neq s}
\delta_{i_s i_k}  L_{i_1}\cdots \hat{L}_{i_s}\cdots\hat{L}_{i_k}\cdots
L_{i_n}.
\end{eqnarray*}
A straightforward calculation shows that the right-hand side of the
last equality is equal to $\Delta (L_{i_1}\cdots L_{i_n})$ and
therefore statement a) is true.

The following identity is easily established by induction on $k$:
\begin{equation}
  \label{identity}
\Delta^k (L_i F) = L_i \Delta^k (F) + 2k \Delta^{k-1}{\partial
  F\over\partial L_i},\quad \forall F\in N, k\geq 1, 1\leq i\leq 3.
\end{equation}
Statement~b) is clearly true for $m=0$. Suppose that it is valid
for $m=k-1$, $k\geq1$, then by using identity (\ref{identity}), we
find that
$$
\Delta^k (L_{i_1}\cdots L_{i_n}) = L_{i_1} \Delta^k
(L_{i_2}\cdots L_{i_n}) + 2k \Delta^{k-1}{\partial\over\partial L_{i_1}}
(L_{i_2}\cdots L_{i_n}).
$$
Summing  over cyclic permutations on both sides of the last equation
and using statement~a) of the Lemma, gives
$$
n \Delta^k (L_{i_1}\cdots L_{i_n}) =
\sum_{(i_1,\ldots,i_n)}L_{i_1} \Delta^k (L_{i_2}\cdots L_{i_n})
+ 2k \Delta^k(L_{i_1}\cdots L_{i_n}),
$$
which shows that statement~b) is true for $m=k$, hence statement~b) holds.

In order to show c),
we apply identity (\ref{identity})
to the Poisson bracket $P(L_i,\Delta^m F)= \sum_{1\leq j,k \leq 3}
\varepsilon_{ijk} L_k \Delta^m {\partial F \over\partial L_{j}}$
and find that
$$
\Delta^m (P(L_i,F)) =\sum_{1\leq j,k \leq3}
\varepsilon_{ijk}L_k {\partial\over\partial L_{j}}\Delta^m  F
+2m \sum_{1\leq j,k\leq 3}
\varepsilon_{ijk} \Delta^m {\partial^2 F \over\partial L_j\partial L_k}.
$$
The first term on the right-hand side is simply $P(L_i,\Delta^m F)$,
while the second term vanishes due to the skew-symmetry of
$\varepsilon_{ijk}$, hence we have found that $\Delta^m (P(L_i,F)) =
P(L_i,\Delta^m F)$, $\forall F\in N, m\geq0$. Setting $F=L_{i_2}\cdots
L_{i_n}$ in this  relation and summing over cyclic permutations yield
$$
\sum_{(i_1,\ldots,i_n)}P(L_{i_1},\Delta^m (L_{i_2}\cdots L_{i_n}))=
\Delta^m( \sum_{(i_1,\ldots,i_n)}P(L_{i_1},
L_{i_2}\cdots L_{i_n})),
$$
whose right-hand side vanishes due to Leibniz property and
skew-symmetry of the Poisson bracket. This shows statement~c).

The differential operator ${\cal D} =\sum_{1\leq k \leq 3} L_k
{\partial\over\partial L_k}$ acting on a homogeneous polynomial $F$ of
degree $\deg(F)$ gives ${\cal D}(F)=\deg(F) F$. The second cochain
$C_2$ in Eq.~(\ref{lproduct}) can be written as
$$
C_2(L_{i},F) = (2+ {\cal D})({\partial F\over\partial L_{i}}).
$$
Let $F=\Delta^m (L_{i_2}\cdots L_{i_n})$ in the preceding equation and
sum over cyclic permutations; this gives
$$
\sum_{(i_1,\ldots,i_n)}C_2(L_{i_1},\Delta^m (L_{i_2}\cdots
  L_{i_n}))= (2+ {\cal D})\Delta^m\Bigl(\sum_{(i_1,\ldots,i_n)}
{\partial\over\partial L_{i_1}} (L_{i_2}\cdots L_{i_n})\Bigr).
$$
The right-hand side of this equation is equal to $(2+ {\cal D})
\Delta^{m+1}(L_{i_1}\cdots L_{i_n})$ by statement~a) of the Lemma;
since $\Delta^{m+1}(L_{i_1}\cdots L_{i_n})$ is a homogeneous
polynomial of degree $n-2m+2$, we have
$(2+ {\cal D})\Delta^{m+1}(L_{i_1}\cdots L_{i_n})
= (n-2m)\Delta^{m+1}(L_{i_1}\cdots L_{i_n})$, and this shows statement~d).
\end{proof}
Using this technical lemma, we are now in position to prove the
central result of this section.
\begin{proposition}
Let $H_n$ be the subspace of $N$ consisting of homogeneous
polynomials of degree~$n$. There exist constants $a(n,r)$,
$n, r\geq 0$, such that the product $\aa$ can be written
$$
F\aa G = \sum_{r\geq 0} \nu^{2r} \eta_r(F G), \quad \forall F,G\in N,
$$
where $\eta_0$ is the identity and $\eta_r\colon N\rightarrow N$,
$r\geq 1$, are linear maps whose restrictions to $H_n$ are given by
$$
\eta_r|_{H_n} = a(n,r) \Delta^r, \quad \forall n,r\geq0.
$$
\end{proposition}
\begin{proof}
In  Section~2, we saw that the product $\aa$ can be written in the form
$$
F\aa G =\sum_{r\geq0} \nu^r \rho_r (FG), \quad\forall F,G\in N,
$$
where $\rho_0= Id$ and for $r\geq1$, $\rho_r$ is a linear
map on $N$.
Consider the product $L_{i_1}\aa \cdots \aa L_{i_n}$ for any
$i_k\in\{1,2,3\}$, $\forall 1\leq k\leq n$. By definition, it is given
by
$$
L_{i_1}\aa \cdots \aa L_{i_n}={1\over n!}\sum_{\sigma\in S_n}
L_{i_{\sigma_1}}\ast\cdots \ast L_{i_{\sigma_n}},
$$
which can be expressed as
\begin{equation}\label{homo3}
L_{i_1}\aa \cdots \aa L_{i_n}={1\over n}\sum_{(i_1,\ldots,i_n)}
L_{i_{1}}\ast(L_{i_{2}}\aa\cdots \aa L_{i_{n}}),
\end{equation}
where $\displaystyle \sum_{(i_1,\ldots,i_n)}$ denotes sum over cyclic
permutations. Using the form (\ref{homo3}) for the product $\aa$
and taking into account that $L_i\ast F= L_i F +\nu P(L_i,F) +
\nu^2 C_2(L_i, F)$, $\forall F\in N$, $1\leq i\leq3$, we find that the
coefficient of the $r$-th power of $\nu$ in $L_{i_1}\aa \cdots \aa
L_{i_n}$ satisfies the following
induction relation for $k\geq2$ and
$n\geq1$ (when $n=1$, we set $L_{i_2}\cdots L_{i_n}=1$):
\begin{eqnarray}\label{rec2}
&&\rho_k(L_{i_1}\cdots L_{i_n}) =\\
&&\nonumber\\
&&\quad {1\over n}\sum_{(i_1,\ldots,i_n)}
\Bigl( L_{i_1}\rho_k(L_{i_2}\cdots L_{i_n})
+
P(L_{i_1},\rho_{k-1}(L_{i_2}\cdots L_{i_n}))
+
C_2(L_{i_1},\rho_{k-2}(L_{i_2}\cdots L_{i_n}))\Bigr),\nonumber
\end{eqnarray}
and for $k=1$ and $n\geq1$:
\begin{equation}\label{rec3}
\rho_1(L_{i_1}\cdots L_{i_n}) =
{1\over n}\sum_{(i_1,\ldots,i_n)}
\Bigl(L_{i_1}\rho_1(L_{i_2}\cdots L_{i_n})
+
P(L_{i_1},L_{i_2}\cdots L_{i_n})\Bigr).
\end{equation}
The second term on the right-hand side of Eq.~(\ref{rec3})
vanishes by skew-symmetry of the Poisson bracket. From $1\aa1=1$,
we have $\rho_1(1)=0$. By
induction on $n$, we easily find that $\rho_1(L_{i_1}\cdots L_{i_n})
=0$, $\forall n\geq 0$,
i.e., $\rho_1=0$.

Let $Z_k$, $k\geq 0$, be the following property:
$$
\rho_{2k}|_{H_n} = a(n,k) \Delta^k,\quad \forall n\geq0,\ {\rm and}\
\rho_{2k+1}=0, \eqno{(Z_k)}
$$
where $a(n,k)$, $n,k\geq 0$, is a constant (notice that for $n<2k$,
$\Delta^k$ vanishes on $H_n$, and in that case the constant $a(n,k)$
can be arbitrary). The
result will be proved by induction on~$k$.

First, remark that $Z_0$ is true with $a(n,0)=1$, $\forall n\geq0$.
Suppose that $Z_k$ is true for $0\leq k\leq r-1$, for some $r\geq 1$.
By hypothesis $\rho_{2r-1}=0$ and $\rho_{2r-2}|_{H_n}=a(n,r-1)\Delta^{r-1}$,
so the induction relation (\ref{rec2}) takes the form
\begin{eqnarray}
&&\rho_{2r}(L_{i_1}\cdots L_{i_n}) =\nonumber\\
&&\nonumber\\
&&\quad{1\over n}\sum_{(i_1,\ldots,i_n)}\Bigl(
 L_{i_1}\rho_{2r}(L_{i_2}\cdots L_{i_n})
+ {a(n-1,r-1)}
C_2(L_{i_1},\Delta^{r-1}(L_{i_2}\cdots L_{i_n}))\Bigr),
\quad \forall n\geq1,\nonumber
\end{eqnarray}
and by application of statement d) of Lemma~2 to the second term
on the right-hand side, we find that
\begin{eqnarray}\label{recrr}
&&\rho_{2r}(L_{i_1}\cdots L_{i_n}) =\\
&&\nonumber\\
&&{1\over n}\Bigl(\sum_{(i_1,\ldots,i_n)}
 L_{i_1}\rho_{2r}(L_{i_2}\cdots L_{i_n})\Bigr)
+ {a(n-1,r-1)(n-2r+2)\over n}
\Delta^{r}(L_{i_1}\cdots L_{i_n}),
\quad \forall n\geq1,\nonumber
\end{eqnarray}
the case $n=0$ is trivially verified by observing that
$\rho_{2r}(1)=0$ for $r\geq1$.

Now we will show by induction on $n$ that any linear map $\gamma$ on $N$
which satisfies along with $\gamma(1)=0$, the following relation
for some given $r\geq1$:
\begin{equation}\label{recrrg}
\gamma(L_{i_1}\cdots L_{i_n}) =
{1\over n}\Bigl(\sum_{(i_1,\ldots,i_n)}
 L_{i_1}\gamma(L_{i_2}\cdots L_{i_n})\Bigr)
+ \alpha_n
\Delta^{r}(L_{i_1}\cdots L_{i_n}),
\quad \forall n\geq1,
\end{equation}
where the $\alpha_n$'s are constants, must be of the form
$\gamma(L_{i_1}\cdots L_{i_n}) =\beta_n \Delta^r(L_{i_1}\cdots
L_{i_n})$, i.e. $\gamma|_{H_n} = \beta_n \Delta^r$ for some constants
$\beta_n$. Suppose $\gamma|_{H_m} = \beta_m\Delta^r$ for $0\leq m\leq
n-1$, then the induction relation (\ref{recrrg}) can be written as
$$
\gamma(L_{i_1}\cdots L_{i_n}) =
{\beta_{n-1}\over n}\Bigl(\sum_{(i_1,\ldots,i_n)}
 L_{i_1}\Delta^{r}(L_{i_2}\cdots L_{i_n})\Bigr)
+ \alpha_n
\Delta^{r}(L_{i_1}\cdots L_{i_n}),
\quad \forall n\geq1,
$$
which by statement b) of Lemma~2 yields
$$
\gamma(L_{i_1}\cdots L_{i_n})=\Bigl((n-2r){\beta_{n-1}\over n}
+ \alpha_n\Bigr)
\Delta^{r}(L_{i_1}\cdots L_{i_n}),\quad \forall n\geq1,$$
and this shows that $\gamma$ must be
proportional to $\Delta^{r}$ on every $H_n$, $n\geq0$, (the case
$n=0$ being trivially verified when $r\geq 1$, since $\Delta^r(1)=0$).

Let us apply the preceding result to the induction relation
(\ref{recrr}); we readily find that $\rho_{2r}(L_{i_1}\cdots L_{i_n})
= a(n,r) \Delta^r (L_{i_1}\cdots L_{i_n})$ and the constant
$a(n,r)$ is given by
\begin{equation}\label{reca}
a(n,r) = {1\over n}\Bigl((n-2r) a(n-1,r) + (n-2r+2)a(n-1,r-1)\Bigr),
\quad \forall n\geq1,
\end{equation}

Hence $\rho_{2r}|_{H_n}=a(n,r)\Delta^r$ and in order to complete the proof
we only need to show that $\rho_{2r+1}=0$.
Under the induction hypothesis we have $\rho_{2r-1}=0$, and we just
have shown that $\rho_{2r}|_{H_n}=a(n,r)\Delta^r$, hence
the induction relation (\ref{rec2}) for $\rho_{2r}$ yields
\begin{eqnarray}\label{recr5}
&&\rho_{2r+1}(L_{i_1}\cdots L_{i_n}) =\nonumber\\
&&\nonumber\\
&&\quad{1\over n}\sum_{(i_1,\ldots,i_n)}\Bigl(
 L_{i_1}\rho_{2r+1}(L_{i_2}\cdots L_{i_n})
+ {a(n-1,r)}
P(L_{i_1},\Delta^{r}(L_{i_2}\cdots L_{i_n}))\Bigr),
\quad \forall n\geq1.\nonumber
\end{eqnarray}
By statement c) of Lemma~2, the second term on the right-hand side
vanishes and we are left with
$$
\rho_{2r+1}(L_{i_1}\cdots L_{i_n}) =
{1\over n}\sum_{(i_1,\ldots,i_n)}
L_{i_1}\rho_{2r+1}(L_{i_2}\cdots L_{i_n}),\quad \forall n\geq1.
$$
We have $\rho_{2r+1}(1)=0$
and by induction on $n$
we easily find that $\rho_{2r+1}(L_{i_1}\cdots L_{i_n})=0$,
$\forall n\geq0$, i.e, $\rho_{2r+1}=0$. This shows that the property
$Z_r$ is true and completes the proof.
\end{proof}
The expression of the constants $a(n,r)$ appearing in the
statement of Proposition~4 (also cf. Eq.~(\ref{reca})) can be written
explicitly with the help of a generating function. Moreover, for
$n\geq 2 r$, the $a(n,r)$'s are polynomials of degree $r$ in
$n$. This fact will allow to find an expression for  the product
$\aa$ in terms of differential operators.
\begin{lemma}
For $n\geq 2r$, the constants $a(n,r)$ of Proposition~4 are given by:
$$
a(n,r)= {1\over n!} \left({\partial \over \partial\alpha}\right)^n
\left.\left((\cos(\alpha))^{-2}(\tan(\alpha))^{n-2r}\right)
\right|_{\alpha=0}, \quad
n\geq 2r.
$$
Moreover there exist polynomials $p_r$, $r\geq0$, of degree $r$ such that for
$n\geq 2r$ we have $a(n,r)=p_r(n)$.
\end{lemma}
\begin{proof}
Let $X$ be any generator of $\frak{su}(2)$. The
$\ast$-powers and the $\aa$-powers of $X$ coincide and the
$\ast$-exponential \cite{BFFLSI,BFFLSII} and the
$\aa$-exponential \cite{DFST}
of $X$ are identical. Here we fix the deformation parameter $\nu$ to be
$i\hbar/2$, where $\hbar$ is real. The $\ast$-exponential of $X$ has
been explicitly computed in \cite{BFFLSII}:
\begin{equation}\label{expstar}
\exp_\ast \left({tX\over i\hbar}\right)\equiv
\sum_{n\geq0}{1\over n!}({t\over i \hbar})^n X^{n\atop \ast} = \cos^{-2}(t/2)
\exp( {2X\over i \hbar}\tan(t/2)),
\end{equation}
for $t$ in a neighborhood of the origin. By Proposition~4 we have
(we denote ${{\scriptstyle\odot}}_{i\hbar/2}$ by ${{\scriptstyle\odot}}$)
\begin{equation}\label{power}
X^{n\atop {{\scriptscriptstyle\odot}}} =
 \sum_{r=0}^{[n/2]}({i\hbar\over2})^{2r}
a(n,r) \Delta^r X^n = \sum_{r=0}^{[n/2]}({i\hbar\over2})^{2r}a(n,r)
{n!\over (n-2r)!}X^{n-2r},
\end{equation}
where $[n/2]$ denotes the integer part of $n/2$.  The function
$\phi(\alpha,\beta)= \cos^{-2}(\alpha)
\exp(\beta\tan(\alpha))$ is analytic in a neighborhood of the origin
of ${\Bbb C}^2$ and notice that $\exp_\ast \left({tX\over i\hbar}\right)
=\exp_{{\scriptscriptstyle\odot}} \left({tX\over i\hbar}\right)
=\phi({t\over 2}, {2X\over i\hbar})$. By Eq.~(\ref{power}), we have
for $t$ in a neighborhood of $0$,
$$
\exp_{{\scriptscriptstyle\odot}} \left({tX\over i\hbar}\right)\equiv
\sum_{n\geq0}{1\over n!}({t\over i \hbar})^n X^{n\atop
{{\scriptscriptstyle\odot}}}
= \sum_{n\geq0}\sum_{r=0}^{[n/2]}({t\over i \hbar})^n
({i\hbar\over2})^{2r}{a(n,r)\over (n-2r)!}X^{n-2r}
$$
or
$$
\phi({t\over 2}, {2X\over i\hbar})
=\sum_{n\geq0}\sum_{r=0}^{[n/2]}{a(n,r)\over (n-2r)!}
({t\over2})^n ({2X\over i \hbar})^{n-2r},
$$
so  ${a(n,r)\over (n-2r)!}$, for $n\geq 2r$, appears to be the
coefficient of $\alpha^n \beta^{n-2r}$ in the Taylor series around
the origin of $\phi(\alpha,\beta)$, hence
$$
a(n,r)= {1\over n!} \left({\partial \over \partial\alpha}\right)^n
\left({\partial \over \partial\beta}\right)^{n-2r}
\left.\left((\cos(\alpha))^{-2}(\exp\beta \tan(\alpha))\right)
\right|_{\alpha=\beta=0}, \quad
n\geq 2r,
$$
or
$$
a(n,r)= {1\over n!} \left({\partial \over \partial\alpha}\right)^n
\left.\left((\cos(\alpha))^{-2}(\tan(\alpha))^{n-2r}\right)
\right|_{\alpha=0}, \quad
n\geq 2r.
$$
{}From the last equality, we see also that $a(n,r)$ is the coefficient
of  $\alpha^n$ in the Taylor series of
$\cos^{-2}(\alpha)(\tan(\alpha))^{n-2r}$. Using
the Taylor series expansions for $\cos^{-1}(\alpha)$ and $\tan(\alpha)$:
$$
\cos^{-1}(\alpha)= \sum_{n\geq0}\gamma^{}_{n}
\alpha^{2n},\quad \tan(\alpha)= \sum_{n\geq0}\tau^{}_n \alpha^{2n+1},
$$
where $\gamma^{}_{n}= E_{2n}/(2n)!$ and
$\tau^{}_n = 2^{2n+2}(2^{2n+2}-1) B_{2n+2}/(2n+2)!$, the constants
$E_k$ (resp. $B_k$) being the Euler (resp. Bernoulli) numbers,
we find that:
\begin{equation}\label{euler}
a(n,r) = \sum_{j_1 + \cdots + j_{n-2r+2}=r\atop j_1,\ldots, j_{n-2r+2}\geq0}
\gamma^{}_{j_1}\gamma^{}_{j_2}\tau^{}_{j_3}\cdots
\tau^{}_{j_{n-2r+2}},
\quad n\geq 2r.
\end{equation}
Let $b(s,r)=a(s+2r,r)$, $s,r\geq0$, $A_k=\sum_{i+j=k\atop i,j \geq0}
\gamma^{}_{i}\gamma^{}_{j}$, $k\geq0$, and
\begin{equation}\label{csk}
c(s,k)=
\sum_{j_1 + \cdots + j_{s}=k\atop j_1,\ldots, j_{s}\geq0}
\tau^{}_{j_1}\cdots\tau^{}_{j_{s}}
\quad s\geq1, k\geq0,
\end{equation}
and $c(0,k)= \delta_{0k}$, $k\geq0$. Notice that $\tau^{}_{0}=1$, so
in particular we have $c(s,0)=1, s\geq0$.
{}From Eq.~(\ref{euler}), we find that $b(s,r)=\sum_{k=0}^{r}A_{r-k} c(s,k)
=A_k +\sum_{k=1}^{r}A_{r-k} c(s,k)$.
We will show that the $c(s,k)$'s are polynomials of degree $k$ in $s$.

For any integer $k\geq1$, we denote by $d(k)$ the number of partitions of $k$,
i.e, the number of ways to write $k$ as a sum of strictly positive integers:
$k=n_1+\cdots+n_p$, with $n_1\geq \cdots \geq n_p$.
We call $p$ the length of the partition $(n_1,\ldots,n_p)$.
Obviously we have $1\leq p \leq k$. Let $B(k,p)$ be the set of
partitions of $k$ of length $p$. For a partition $(n_1,\ldots,n_p)$
of $B(k,p)$, let $m_i$ be the multiplicity with which a given integer
$a_i>0$ appears in $(n_1,\ldots,n_p)$, so that $k=m_1 a_1+\cdots+m_ra_r$
and $a_1>\cdots>a_r$. To that partition $(n_1,\ldots,n_p)$
we associate a symmetry factor given by
$$m(n_1,\ldots,n_p)=\frac{1}{m_1!\cdots m_r!}.
$$

Let $s\geq1$ be an integer. Let $j_1,\ldots,j_s \geq 0$ be integers such
that $j_1 + \cdots + j_{s}=k$. We can
associate to the integers $j_1,\ldots,j_s$, by ordering them, a
partition
$(n_1,\ldots,n_p)$
of $k$ for some $p$. Conversely, for  a partition $(n_1,\ldots,n_p)$
of length $p$ of $k$
there are $\frac{s!}{(s-p)!} m(n_1,\ldots,n_p)$
ways to associate a set of non-negative
integers satisfying
$j_1 + \cdots + j_{s}=k$ when $s\geq p$.
Hence the sum in Eq.~(\ref{csk}) for $s\geq1, k\geq1$ can be written as:
$$
c(s,k)=
\sum_{p=1}^{k} s(s-1)\cdots (s-p+1)\sum_{(n_1,\ldots,n_p)\in B(k,p)}
m(n_1,\ldots,n_p)\tau^{}_{n_1}\cdots\tau^{}_{n_{p}},
$$
with an obvious interpretation when $s< k$. Since $c(0,k)=0$ for
$k\geq1$, the preceding formula
also covers the case $s=0$. Also we saw that we have $c(s,0)=1$, $s\geq0$,
hence $c(s,k)$, $k\geq0$, is a polynomial in $s$. The term $p=k$ in
the preceding sum corresponds
to a polynomial of degree $k$ and since it is the polynomial with
maximal degree in that sum,
we conclude that $c(s,k)$ is a polynomial of degree $k$ in $s$. By
retracing the preceding steps, we conclude that there exist polynomials
$p_r$, $r\geq0$, of degree $r$ such that for
$n\geq 2r$ we have $a(n,r)=p_r(n)$. With the previous notations, these
polynomials are given by $p_0(n)=1$ and for $r\geq1$:
\begin{equation}\label{prn}
p_r(n)= A_r + \sum_{p=1}^{r} z_{p,r} (n-2r)\cdots(n-2r-p+1),
\end{equation}
where $z_{p,r} = \sum_{k=p}^r A_{r-k}\sum_{(n_1,\ldots,n_p)\in B(k,p)}
m(n_1,\ldots,n_p)\tau^{}_{n_1}\cdots\tau^{}_{n_{p}}$.
\end{proof}
This lemma allows to express the cochains of the $\aa$-product on
$\frak{su}(2)^*$ by differential operators and
we conclude this section by a theorem which summarizes the
construction developed here.
\begin{theorem}
The $\aa$-product associated with the invariant star-product on
$\frak{su}(2)^* $ (defined by
Eq.~(\ref{lproduct})) admits the following form:
$$
F\aa G = FG + \sum_{r\geq1} \nu^{2r}\eta_r(FG),\quad F,G\in N,
$$
where $\eta_r\colon  N\rightarrow N$, $r\geq1$, are differential
operators given
by
\begin{equation}\label{nr}
\eta_r(F)= \left(A_r + \sum_{p=1}^{r} z_{p,r} {\cal D}({\cal D}-1)
\cdots({\cal D}-p+1)\right)
\Delta^r(F), \quad F\in N,
\end{equation}
where ${\cal D} =\sum_{1\leq k \leq 3} L_k
{\partial\over\partial L_k}$ and where the constants $A_r$ and
$z_{p,r}$ are given in Lemma~3.
This $\aa$-product is strongly (but not weakly) non-trivial and
has a unique extension to $F,G \in C^{\infty}({\Bbb R}^3)$.
\end{theorem}
\begin{proof}
{}From Proposition~4, we have $\eta_r|_{H_n} = a(n,r) \Delta^r$ for
$n,r\geq0$. Notice that $\Delta^r$ vanishes on $H_n$ for $n<2r$ and by
 Lemma~3  we have for $n\geq 2r$,
$a(n,r)=p_r(n)$, where the $p_r$'s are the polynomials defined by
Eq.~(\ref{prn}).
Hence $\eta_r|_{H_n} = p_r(n)\Delta^r$. Let
${\cal D} =\sum_{1\leq k \leq 3} L_k
{\partial\over\partial L_k}$. Since ${\cal D}|_{H_n}=n$ and $\Delta^r$ maps
$H_n$ to $H_{n-2r}$ for $n\geq 2r$ and is $0$ otherwise, it follows
that the restriction of the differential operator $p_r({\cal D} -2r)
\Delta^{2r}$ to $H_n$ coincides
with $\eta_r|_{H_n}$,
$n\geq0$, therefore $\eta_r=p_r({\cal D} -2r) \Delta^{2r}$ on $N$. By
using the explicit expression for $p_r(n)$ given by Eq.~(\ref{prn}),
relation (\ref{nr}) follows.
The last statement of the Theorem follows from the fact that
$\pi(F\aa G)= FG$, $F,G\in C^{\infty}({\Bbb R}^3)$, which implies, as for
the polynomial case, that the $\aa$-product is associative.
\end{proof}
\section{Concluding Remarks}
{\it Spectrality.}
As indicated in \cite{DFST}, in the framework of generalized
deformations, the spectrum of an observable is obtained through its
corresponding ${\scriptstyle\odot}$-exponential (cf. proof of Lemma~3). Consider
the invariant star-product on $\frak{su}(2)^\ast$ (with $\nu=i\hbar/2$),
Eq.~(\ref{expstar}) gives explicitly the $\ast$-exponential of a linear
element $X$, and from that one finds that the $\ast$-spectrum of $X$ is
discrete \cite{BFFLSII}. For $X$ linear, we have noticed that
$\exp_\ast ({tX\over i\hbar})=\exp_{\scriptscriptstyle\odot}({tX\over i\hbar})$, thus
the ${\scriptstyle\odot}$-spectrum and the $\ast$-spectrum of $X$ coincide.
The ${\scriptstyle\odot}$-spectrum of $X$ is discrete, hence non-trivial; although
the $\aa$-product of Section~4 is weakly trivial.
The situation changes drastically if one considers, e.g., the square of
the total angular momentum $H=\frac{1}{2}(L_1^2+L_2^2+L_3^2)$. Then
it is no longer true that the corresponding spectra coincide.

Consider the Zariski product $\bullet$ associated with the invariant
star-product $\frak{su}(2)^\ast$. The observable $H$ is an irreducible
polynomial (over the reals), hence the $\ast$-powers and
$\bullet$-powers of $H$ are identical and the $\bullet$-spectrum of $H$
is nothing but its  $\ast$-spectrum. The situation is similar for the
case of a linear polynomial $X$ on $\frak{su}(2)^\ast$. Hence while
the $\aa$-product does not give the usual spectrum for the square
of the total angular momentum (or the rotational kinetic energy), the
Zariski product does.
\vskip1mm
\noindent{\it Quantized Nambu Bracket.}
A quantization of the classical Nambu bracket is achieved by
replacing the usual product by the $\aa$-product of Section~4.
Due to the properties of the $\aa$-product, it is easy to see
that actually the quantized Nambu bracket is given by:
$$
[F,G,H]_{\aa} = \tilde T\circ \lambda_0( \{F,G,H\}),
\quad F,G,H\in C^\infty({\Bbb R}^3),
$$
where $\{F,G,H\}$ denotes the classical Nambu bracket on ${\Bbb R}^3$,
i.e., the Jacobian. Though the Leibniz rule is
not satisfied for the
$\aa$-product, this quantized Nambu bracket does satisfy the
Fundamental Identity. Indeed only the weaker form of the Leibniz rule
$$
F\aa(\frac{\partial}{\partial L_i}(G\aa H) -
G\aa\frac{\partial H}{\partial L_i} -
\frac{\partial G}{\partial L_i}\aa H)
=\tilde T\circ \lambda_0(F(\frac{\partial}{\partial L_i}(G H) -
G\frac{\partial H}{\partial L_i} -
\frac{\partial G}{\partial L_i}H)) =0,
$$
is required to ensure that the Fundamental Identity is verified by the
quantized Nambu bracket.

Finally let us mention that, as for the $\aa$-product case,
the quantized Nambu bracket is strongly (but not weakly) non-trivial.
\begin{acknowledgement}
The authors are indebted to D.~Sternheimer
for very useful discussions and careful readings of the manuscript.
We also thank  L.~Takhtajan and Ph.~Bonneau for their useful remarks.
G. D. wishes to thank H.~Araki and I.~Ojima for wonderful
hospitality at RIMS.
\end{acknowledgement}


\begin{thebibliography}{99}
\bibitem{BFFLSI} Bayen, F., Flato, M., Fr\o nsdal, C., Lichnerowicz, A.,
Sternheimer, D.:  Deformation Theory and Quantization: I. Deformations of
Symplectic Structures, {\it Ann. Phys.} {\bf 111} (1978), 61--110.
\bibitem{BFFLSII} Bayen, F., Flato, M., Fr\o nsdal, C., Lichnerowicz, A.,
Sternheimer, D.:  Deformation Theory and Quantization: II.
 Physical Applications, {\it Ann. Phys.} {\bf 111} (1978), 111--151.
\bibitem{DFST} Dito, G., Flato, M., Sternheimer, D., Takhtajan, L.:
Deformation Quantization and Nambu Mechanics, to appear in
{\it Commun. Math. Phys.} (1996).
\bibitem{Di} Dito, G.: In preparation.
\bibitem{FF} Flato, M., Fr\o nsdal, C.: Unpublished (1992).
\bibitem{Na} Nambu, Y.:
Generalized Hamiltonian Dynamics. {\it Phys. Rev. D} {\bf 7} (1973),
2405--2412.
\bibitem{Ta} Takhtajan, L.:
 On Foundation of the Generalized Nambu Mechanics, {\it  Commun. Math. Phys.}
{\bf 160} (1994), 295--315.
\end{thebibliography}
\end{document}